\documentclass[12pt]{article}
\usepackage{cite}
\usepackage{CJK}
\usepackage{amssymb}
\usepackage{amsmath}
\usepackage{enumerate}
\usepackage{graphicx}
\usepackage{amsfonts}
\usepackage{url}
\usepackage{bm}
\usepackage[marginal]{footmisc}

\usepackage{pifont}
\usepackage{epsfig,subfigure,dsfont,amsthm,amsbsy,mathrsfs,amscd}

\def\la{\langle}
\def\ra{\rangle}
\def\be{\begin{equation}}
\def\ee{\end{equation}}
\def\ba{\begin{array}}
\def\ea{\end{array}}
\input amssym.def

\newcommand\btd{\raise 2pt \hbox{$\hat\bigtriangledown$}\hskip 1.5pt}
\newcommand\bt{\raise 2pt \hbox{$\bigtriangledown$}\hskip 1.5pt}
\begin{document}
 \title{\large\bf Multi-observable Uncertainty Relations in Product Form of Variances}
 \author{Hui-Hui Qin$^{1}$ \& Shao-Ming Fei$^{2,3}$ $^\ast$\& Xianqing Li-Jost$^{3}$ \\[10pt]
\footnotesize
\small $^1$Department of Mathematics, School of Science, South
China University of Technology,\\ \small Guangzhou 510640, China\\
\footnotesize
\small $^2$School of Mathematical Sciences, Capital Normal University,
Beijing 100048, China\\
\footnotesize
\small $^{3}$Max-Planck-Institute for Mathematics in the Sciences, Leipzig 04103, Germany}
\date{}

\maketitle

\centerline{$^\ast$ Correspondence to feishm@cnu.edu.cn}
\bigskip

\begin{abstract}

We investigate the product form uncertainty relations of variances for $n\,(n\geq 3)$ quantum observables.
In particular, tight uncertainty relations satisfied by three observables has been derived, which is shown
to be better than the ones derived from the strengthened Heisenberg and the generalized Schr\"{o}dinger
uncertainty relations, and some existing uncertainty relation for three spin-half operators.
Uncertainty relation of arbitrary number of observables is also derived.
As an example, the uncertainty relation satisfied by the eight Gell-Mann matrices is presented.
\end{abstract}
\bigskip

Uncertainty relations \cite{Heisenberg} are of profound significance in quantum mechanics and also in quantum information theory
like quantum separability criteria and entanglement detection \cite{Guehne,Guehne',Hofmann}, security analysis of quantum key distribution
in quantum cryptography \cite{Fuchs}, and nonlocality \cite{Oppenheim}. The Heisenberg-Robertson uncertainty relation \cite{Heisenberg,Robertson,Robertson'}
presents a lower bound on the product of the standard deviations of two observables, and
provides a trade-off relation of measurement errors of these two observables for any given quantum states.
Since then different types of uncertainty relations have been studied. There are many ways to quantify the uncertainty of measurement outcomes. In
\cite{Heisenberg,Robertson,Robertson',Trifonov,Ivan,Schroedinger,Serafini,Massar,Kechrimparis} the product
uncertainty relations for the standard deviations of the distributions of observables is studied.
In \cite{Pati,Maccone,Chen} the uncertainty relations related to the sum of varinces
or standard deviations have been investigated. And in
\cite{Deutsch,Maassen,Friedland,SR,CP,13,14,15,16,pz1,pz2,r15,y9} entropic uncertainty relations
with majorization technique are explored. Uncertainty relations are also described in terms of the noise and disturbance \cite{y6,BLW1},
and according to successive measurements \cite{y7,y8,y16,y17}.
Let $\rho$ be a quantum state and $A$ be a quantum mechanical observable.
The variance of $A$ with respect to the state $\rho$ is defined by
$(\Delta A)^{2}=\la A^{2}\ra-\la A\ra^{2}$, where
$\la A\ra=tr(A\rho)$ is the mean value of $A$. 
From Heisenberg and Robertson \cite{Heisenberg,Robertson},
the product form uncertainty relation of two observables $A$ and $B$ is expressed as
\be\label{eq1}
(\Delta A)^{2}(\Delta B)^{2} \geq \frac{1}{4}|\la [A, B]\ra|^{2},
\ee
which is further improved by Schr\"{o}dinger,
\be\label{eq2}
(\Delta A)^{2}(\Delta B)^{2}\geq \frac{1}{4}|\la [A, B]\ra|^{2}
+\frac{1}{4}|\la \{A,B\}\ra-\la A \ra\la B \ra|^{2},
\ee
where $\{A,B\}$ is the anticommutator of $A$ and $B$.

However, till now one has no product form
uncertainty relations for more than two observables.
Since there is no relations like Schwartz inequality
for three or more objects, generally it is difficult to have
a nontrivial inequality satisfied by the quantity $(\Delta A)^{2}(\Delta B)^{2}...(\Delta C)^{2}$.
In \cite{Kechrimparis} Kechrimparis and Weigert
obtained a tight product form uncertainty relation for three canonical observables $\hat{p}$, $\hat{q}$
and $\hat{r}$,
\be\label{eq9}
(\Delta \hat{p})^{2} (\Delta  \hat{q})^{2} (\Delta \hat{r})^{2} \geq (\tau \frac{\hbar}{2})^{3},
\ee
where $\tau=\frac{2}{\sqrt{3}}$, $\hat{q}$ and $\hat{p}$ are the position and momentum respectively,
and $\hat{r}=-\hat{p}-\hat{q}$. As $\tau>1$ the relation (\ref{eq9}) is stronger than the one obtained directly
from the commutation relations $[\hat{p},\hat{q}]=[\hat{q},\hat{r}]=[\hat{r},\hat{p}]=\frac{\hbar}{i}$ and
the uncertainty relation (\ref{eq1}).
Here the `observable' $\hat{r}=-\hat{p}-\hat{q}$ is not a physical quantity, neither independent in this triple. In fact,
besides the dual observables like position and momentum, there are also triple physical observables like spin, 
isospin (isotopic spin) related to the strong interaction in particle physics, angular momentum that their components are pairwise noncommutative.

Generally speaking, uncertainty relations are equalities or inequalities satisfied by functions such as
polynomials of the variances of a set of observables.
In this paper, we investigate the product form uncertainty relations of multiple observables.
We present a new uncertainty relation which gives better characterization of the uncertainty
of variances.

\medskip
\noindent{\bf Results}
\medskip

{\bf Theorem 1}~The product form uncertainty of three observables $A$, $B$, $C$ satisfies the following relation,
\be\label{eq3}
\begin{aligned}
\displaystyle & (\Delta A)^{2}(\Delta B)^{2}(\Delta C)^{2} \geq \\
\displaystyle & (\Delta A)^{2}|\la BC \ra-\la B\ra \la C\ra|^{2} +
(\Delta B)^{2}|\la CA\ra-\la C\ra \la A\ra|^{2}
+(\Delta C)^{2}|\la AB\ra-\la A\ra\la B\ra|^{2}\\
\displaystyle &\quad -
2Re\{(\la AB\ra- \la A\ra\la B\ra)
(\la BC\ra-\la B\ra \la C\ra)
(\la CA \ra-\la C\ra \la A\ra)\},
\end{aligned}
\ee
where $Re\{S\}$ stands for the real part of $S$.

See Methods for the proof of Theorem 1.

The right hand side of (\ref{eq3}) contains terms like $\la BC \ra$ and $\la CA \ra$.
These terms can be expressed in terms of the usual form of commutators and anti-commutators.
From the Hermitianity of observables and $(\la AB \ra)^{*}=\la BA \ra$,
one has $\la AB \ra=\frac{1}{2}(\la [A,B]\ra + \la \{A,B\}\ra)$.
By using these relations formula (\ref{eq3}) can be reexpressed as,
\be\label{eq5}
\begin{aligned}
\displaystyle (\Delta A)^{2}(\Delta B)^{2}(\Delta C)^{2}
&\geq (\la A^{2} \ra-\la A \ra^{2})
(\frac{1}{4}|\la  [B,C] \ra|^{2}+|\frac{1}{2}\la  \{B,C\} \ra-\la B  \ra\la C \ra|^{2})\\
 \displaystyle &\quad+(\la B^{2} \ra-\la B \ra^{2})
(\frac{1}{4}|\la  [C,A] \ra|^{2}+|\frac{1}{2}\la  \{C,A\} \ra-\la C  \ra\la A \ra|^{2}) \\
\displaystyle &\quad+(\la C^{2} \ra-\la C \ra^{2})
(\frac{1}{4}|\la  [A,B] \ra|^{2}+|\frac{1}{2}\la  \{A,B\} \ra-\la A  \ra\la B \ra|^{2}) \\
\displaystyle &\quad -\frac{1}{4}(\la \{A,B\} \ra-2\la A  \ra\la B  \ra)
(\la\{B,C\} \ra-2\la B  \ra\la C \ra)
(\la \{C,A\} \ra-2\la C  \ra\la A \ra)\\
\displaystyle &\quad-\frac{1}{4}(\la \{A,B\} \ra-2\la A  \ra\la B  \ra)
\la [B,C] \ra\la [C,A] \ra\\
\displaystyle &\quad-\frac{1}{4}\la [A,B] \ra
(\la \{B,C\} \ra-2\la B  \ra\la C  \ra)\la [C,A] \ra\\
\displaystyle &\quad-\frac{1}{4}(\la [A,B] \ra)
\la [B,C] \ra(\la \{C,A\} \ra-2\la C  \ra\la A  \ra).
\end{aligned}
\ee

Formulae (\ref{eq3}) or (\ref{eq5}) give a general relation satisfied by $(\Delta A)^{2}$, $(\Delta B)^{2}$ and $(\Delta C)^{2}$.
To show the advantages of this uncertainty inequality, let us consider the case of
three Pauli matrices $A=\sigma_{x}$, $B=\sigma_{y}$, and $C=\sigma_{z}$.
Our Theorem says that
\be\label{eq6}
\begin{aligned}
\displaystyle & (\Delta \sigma_{x})^{2}(\Delta \sigma_{y})^{2}(\Delta \sigma_{z})^{2} \\
\displaystyle &\quad \geq  (1-\langle \sigma_{x} \rangle^{2})
(|\la \sigma_{x}\ra|^{2}+|\la \sigma_{y} \ra\la\sigma_{z}\ra|^{2})\\
\displaystyle &\quad+(1-\la \sigma_{y} \ra^{2})
(|\la \sigma_{y}\ra|^{2}+|\la \sigma_{z} \ra\la \sigma_{x} \ra|^{2}) \\
\displaystyle &\quad+(1-\la\sigma_{z} \ra^{2})
(|\la \sigma_{z}\ra|^{2}+|\la\sigma_{x} \ra\la \sigma_{y}  \ra|^{2})
+2\la \sigma_{x}  \ra^{2}\la \sigma_{y} \ra^{2}\la \sigma_{z}  \ra^{2}\\
\displaystyle &\quad-2\la \sigma_{x}  \ra^{2}\la \sigma_{y}  \ra^{2}
-2\la \sigma_{y} \ra^{2}\la \sigma_{z} \ra^{2}
-2\la \sigma_{z} \ra^{2}\la \sigma_{x} \ra^{2}.
\end{aligned}
\ee
Let the qubit state $\rho$ to be measured be given in the Bloch representation with Bloch vector
$\overrightarrow{r}=(r_{1},r_{2},r_{3})$, i.e.
$\rho=\frac{1}{2}(I+\overrightarrow{r}\cdot\overrightarrow{\sigma})$,
where $\overrightarrow{\sigma}=(\sigma_{x},\sigma_{y},\sigma_{z})$, $\sum^{3}_{i=1}r^{2}_{i} \leq 1$.
Then one has
$(\Delta \sigma_{x})^{2}(\Delta \sigma_{y})^{2}(\Delta \sigma_{z})^{2}
=(1-r^{2}_{1})(1-r^{2}_{2})(1-r^{2}_{3})$.
And the uncertainty relation (\ref{eq6}) has the form
\be \label{eq7}
\begin{aligned}
\displaystyle & (\Delta \sigma_{x})^{2}(\Delta \sigma_{y})^{2}(\Delta \sigma_{z})^{2}=(1-r^{2}_{1})(1-r^{2}_{2})(1-r^{2}_{3}) \\
\displaystyle &\geq \quad \sum^{3}_{i=1}r^{2}_{i}- \sum^{3}_{i=1}r^{4}_{i}-\sum_{1\leq i<j \leq3}r^{2}_{i}r^{2}_{j}
-r^{2}_{1}r^{2}_{2}r^{2}_{3}.
\end{aligned}
\ee
The difference between the right and left hand side of (\ref{eq7}) is
$(1-\sum^{3}_{i=1}r^{2}_{i})^{2}$. That is, the equality holds iff $\sum^{3}_{i=1}r^{2}_{i}=1$.
Therefore, the uncertainty inequality is tight for all pure states. 
Usually, a lower bound on the product of variances implies a lower bound on the sum of variances 
\cite{PRA86024101}.
Indeed in these cases the lower bound in (\ref{eq7}) also gives
a tight lower bound of the sum of variances, since 
$(\Delta \sigma_{x})^{2}+(\Delta \sigma_{y})^{2}+(\Delta \sigma_{z})^{2}=
3-\sum^{3}_{i=1}r^{2}_{i}=2
\geq 3\sqrt[3]{(\Delta \sigma_{x})^{2}(\Delta \sigma_{y})^{2}(\Delta \sigma_{z})^{2}} \geq 3\sqrt[3]{\mathcal{L}_{\ref{eq7}}}=2$,
where $\mathcal{L}_{\ref{eq7}}$ is the right hand side of (\ref{eq7}).

In fact, from the Heisenberg and Robertson uncertainty relation,
One has $(\Delta \sigma_{x})^{2} (\Delta \sigma_{y})^{2} (\Delta \sigma_{z})^{2} \geq
|\frac{\la[\sigma_{x},\sigma_{y}]\ra}{2}| |\frac{\la[\sigma_{y},\sigma_{z}]\ra}{2}||\frac{\la[\sigma_{z},\sigma_{x}]\ra}{2}|$.
However, this inequality is not tight.
In \cite{Bin} the inequality is made tight by multiplying a constant factor $\tau=\frac{8}{3\sqrt{3}}$ on the right hand side,
and the tighten uncertainty relation reads,
\be\label{eq10}
(\Delta\sigma_{x})^{2} (\Delta\sigma_{y})^{2} (\Delta\sigma_{z})^{2} \geq \frac{8}{3\sqrt{3}}|r_{1}r_{2}r_{3}|.
\ee

Let us compare the lower bound of (\ref{eq7}) with that of (\ref{eq10}).
The difference of these two bounds satisfies the following inequality,
\be\nonumber
\begin{aligned}
\displaystyle &\left(\sum^{3}_{i=1}r^{2}_{i}-
\sum^{3}_{i=1}r^{4}_{i}-\sum_{1\leq i<j\leq 3}r^{2}_{i}r^{2}_{j}
-r^{2}_{1}r^{2}_{2}r^{2}_{3}\right)-\frac{8}{3\sqrt{3}}|r_{1}r_{2}r_{3}|\\
\displaystyle &\quad \geq
\sum_{1\leq i<j \leq 3}r^{2}_{i}r^{2}_{j}-r^{2}_{1}r^{2}_{2}r^{2}_{3}-
\frac{8}{3\sqrt{3}}|r_{1}r_{2}r_{3}| \\
\displaystyle &\quad \geq
3(r_{1}r_{2}r_{3})^{\frac{4}{3}}-(r_{1}r_{2}r_{3})^{2}-\frac{8}{3\sqrt{3}}|r_{1}r_{2}r_{3}|\geq 0
\end{aligned}
\ee
for all $r_{1}r_{2}r_{3} \in [-1,1]$, where in the first inequality we have used the fact that
$\|\overrightarrow{r}\|^{2}=\sum^{3}_{i=1}r^{2}_{i} \leq 1$.
This illustrates that the uncertainty relation of three Pauli operators from Theorem 1 is stronger
than the tighten uncertainty relation (\ref{eq10}), obtained from the Heisenberg and Robertson uncertainty relation.

From the generalized Schr\"{o}dinger uncertainty relation (\ref{eq2}),
one can also get an uncertainty relation for three observables,
\be\label{eq11}
\begin{aligned}
\displaystyle &(\Delta A)^{2}(\Delta B)^{2}(\Delta C)^{2}\\
\displaystyle &\quad \geq
 |\la AB\ra-\la A\ra \la B\ra| |\la BC\ra-\la B\ra \la C\ra|
 |\la CA \ra-\la C\ra \la A \ra|\\
 \displaystyle &\quad =|\la \mathbf{A}\mathbf{B}\ra||\la \mathbf{B}\mathbf{C}\ra|
 |\la \mathbf{C}\mathbf{A}\ra|,
\end{aligned}
\ee
where $|\la XY\ra|^{2}=|\frac{1}{2}\la[X,Y]\ra|^{2}+
|\frac{1}{2}\la \{X,Y\}\ra-\la X\ra\la Y\ra|^{2}$ for $X,Y=A, B, C$, and $\mathbf{A}, \mathbf{B}$ and $\mathbf{C}$
are the variance operators of $A$, $B$ and $C$, respectively, defined by
$\mathbf{O}=O-\la O \ra I$ for any operator $O$.
Comparing directly the right hand side of (\ref{eq4}) with the right hand side of (\ref{eq11}), we obtain
\be
\begin{aligned}
\displaystyle &(\Delta\mathbf{A})^{2}|\la\mathbf{B}\mathbf{C} \ra|^{2} +
(\Delta \mathbf{B})^{2}|\la\mathbf{C}\mathbf{A}\ra|^{2}
+(\Delta \mathbf{C})^{2}|\la\mathbf{A}\mathbf{B}\ra|^{2}\\
\displaystyle &\quad -
2Re\{\la \mathbf{A}\mathbf{B}\ra \la \mathbf{B}\mathbf{C}\ra \la\mathbf{C}\mathbf{A} \ra\}
-|\la \mathbf{A}\mathbf{B}\ra \la \mathbf{B}\mathbf{C}\ra \la\mathbf{C}\mathbf{A} \ra|\\
\displaystyle &\quad \geq 3((\Delta\mathbf{A})^{2}(\Delta\mathbf{B})^{2}(\Delta\mathbf{C})^{2}|\la \mathbf{A}\mathbf{B}\ra
\la \mathbf{B}\mathbf{C}\ra \la\mathbf{C}\mathbf{A} \ra|^{2})^{\frac{1}{3}}\\
\displaystyle &\quad -
2Re\{\la \mathbf{A}\mathbf{B}\ra \la \mathbf{B}\mathbf{C}\ra \la\mathbf{C}\mathbf{A} \ra\}
-|\la \mathbf{A}\mathbf{B}\ra \la \mathbf{B}\mathbf{C}\ra \la\mathbf{C}\mathbf{A} \ra|\\
\displaystyle &\quad \geq
2|\la \mathbf{A}\mathbf{B}\ra \la \mathbf{B}\mathbf{C}\ra \la\mathbf{C}\mathbf{A} \ra|
-2Re\{\la \mathbf{A}\mathbf{B}\ra \la \mathbf{B}\mathbf{C}\ra \la\mathbf{C}\mathbf{A} \ra\}\geq 0,
\end{aligned}
\ee
where the second inequality is obtained by (\ref{eq11}). Hence our uncertainty relation is also stronger than
the one obtained from the generalized Schr\"{o}dinger uncertainty relation.

\begin{figure}[htpb]
\centering\
\includegraphics[width=6.5 cm]{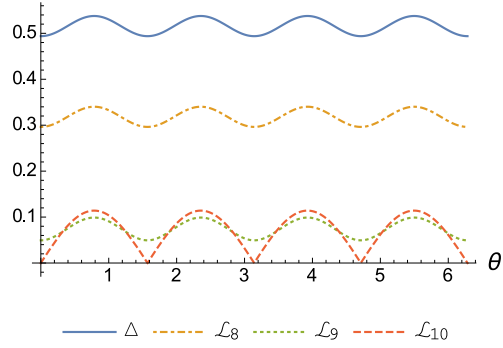}
\caption{{\small (Color online) Uncertainty relations satisfied by observables $\sigma_{x}$, $\sigma_{y}$ and $\sigma_{z}$ with state $\rho$ parameterized by
Bloch vector $\{\frac{1}{3},\frac{2}{3}\cos\theta,\frac{2}{3}\sin\theta\}$:
solid line for $(\Delta\sigma_{x})^{2}(\Delta\sigma_{y})^{2}(\Delta\sigma_{z})^{2}$, dot-dashed line for
lower bound in (\ref{eq7}), dashed line for the lower bound in (\ref{eq10}), dotted line for
lower bound in (\ref{eq11}).}}\label{Fig.1}
\end{figure}

As an example, let us take the Bloch vector of the state $\rho$ to be
$\overrightarrow{r}=\{\frac{1}{3},\frac{2}{3}\cos\alpha,\frac{2}{3}\sin\alpha\}$.
Then we get $(\Delta\sigma_{x})^{2}(\Delta\sigma_{y})^{2}(\Delta\sigma_{z})^{2} \geq
\mathcal{L}_{\ref{eq7}} \geq \max\{\mathcal{L}_{\ref{eq10}},\mathcal{L}_{\ref{eq11}}\}$, where $\mathcal{L}_{\ref{eq7}}$,
$\mathcal{L}_{\ref{eq10}}$ and $\mathcal{L}_{\ref{eq11}}$ are the right hand sides of inequalities (\ref{eq7}),
(\ref{eq10}) and (\ref{eq11}), respectively, see Fig. \ref{Fig.1}.

\medskip
We have presented a product form uncertainty relation for three observables. Our approach can be also used
to derive product form uncertainty relations for multiple observables.
Consider n observables $\{A_{i}\}^{n}_{i=1}$. Denote $I=I_{n}=\{1, 2, \dots, n\}$,
$I_{k}=\{i_{1},i_{2},\dots,i_{k}\}\subseteq I$ with $k$ elements of $I$, $k=1, 2,\dots, n$,
$\overline{I}_{k}=\{i_{k+1},i_{k+2},\dots,i_{n}\}=I\setminus I_{k}$.
Let $\mathcal{I}$ be the set consisting of all the subsets of $I$, and $\mathcal{I}_{k}$ the set
consisting of the subsets of $I$ with $k$ elements. Then we have $\cup I_{k}=\mathcal{I}_{k}$,
$\cup^{n}_{k=1} \mathcal{I}_{k}=\mathcal{I}$. We have

{\bf Theorem 2}~
\begin{equation}\label{T2}
\prod^{n}_{i=1}{(\Delta A_{i})^{2}}\geq \sum^{n}_{k=1}(-1)^{n-k}g(k),
\end{equation}
where
\begin{equation}\nonumber
\begin{aligned}
g(k)=&\displaystyle
(-1)^{n-k}\sum_{I_{k}\in \mathcal{I}_{k}}\prod_{i_{s}\in I_{k}}\Delta A^{2}_{i_{s}}
\{\sum_{i_{k+1},\dots, i_{n} \in \overline{I}_{k}}
\frac{1}{n-k}E_{i_{k+1}i_{k+2}\dots i_{n}}\\
&\displaystyle +
\sum_{i_{k+1},i_{k+2}\in \overline{I}_{k}}(-\frac{1}{2})E_{i_{k+1}i_{k+2}}\sum_{i_{k+3},\dots,i_{n}\in
\overline{I}_{k}\setminus\{i_{k+1},i_{k+2}\}}\frac{1}{n-k-2}E_{i_{k+3}i_{k+4}\dots i_{n}}\\
&\displaystyle +\dots+
\sum_{i_{k+1},\dots,i_{n}\in \overline{I}_{k}}
(-\frac{1}{2})^{\frac{n-k-2}{2}}E_{i_{k+1}i_{k+2}}E_{i_{k+3}i_{k+4}}\dots E_{i_{n-1}i_{n}}\}
\end{aligned}
\end{equation}

when $n-k$ is even, and

\begin{equation}\nonumber
\begin{aligned}
g(k)=
\displaystyle &
(-1)^{n-k}\sum_{I_{k}\in \mathcal{I}_{k}}\prod_{i_{s}\in I_{k}}\Delta A^{2}_{i_{s}}
\{\sum_{i_{k+1},i_{k+2},\dots, i_{n} \in \overline{I}_{k}}
\frac{1}{n-k}E_{i_{k}i_{k+1}i_{n}}\\
\displaystyle & +
\sum_{i_{k+},i_{k+2}\in \overline{I}_{k}}(-\frac{1}{2})E_{i_{k+1}i_{k+2}}
\sum_{i_{k+3},\dots,i_{n}\in
\overline{I}_{k}\setminus\{i_{k+1},i_{k+2}\}}\frac{1}{n-k-2}E_{i_{k+3}i_{k+4}\dots i_{n}}\\
\displaystyle& +\dots+
\sum_{i_{k+1},\dots,i_{n-3}\in \overline{I}_{k}}
(-\frac{1}{2})^{\frac{n-k-3}{2}}E_{i_{k+1}i_{k+2}}\dots E_{i_{n-4}i_{n-3}}
\sum_{i_{n-2},i_{n-1},i_{n}}\frac{1}{3}E_{i_{n-2}i_{n-1}i_{n}}\}
\end{aligned}
\end{equation}

when $n-k$ is odd,$E_{i_{1},\dots, i_{k}}=\la \mathbf{A}_{i_{1}}\mathbf{A}_{i_{2}}\ra\la \mathbf{A}_{i_{2}}\mathbf{A}_{i_{3}}\ra\dots
\la \mathbf{A}_{i_{k-1}}\mathbf{A}_{i_{k}}\ra \la \mathbf{A}_{i_{k}}\mathbf{A}_{i_{1}}\ra$, $\mathbf{A}_{i}s$
are the variance operators of $A_{i}s$.

For instance, we calculate the product form uncertainty relation for the
eight Gell-Mann matrices $\{\lambda_{n}\}^{8}_{n=1}$,
\be\nonumber
\lambda_{1}=
\begin{pmatrix}
0&1&0\\ 1&0&0\\ 0&0&0
 \end{pmatrix},~~
 \lambda_{2}=
 \begin{pmatrix}
 0&-i&0\\ i&0&0\\ 0&0&0
 \end{pmatrix},~~
 \lambda_{3}=
 \begin{pmatrix}
 1&0&0\\ 0&-1&0\\ 0&0&0
 \end{pmatrix},~~
 \lambda_{4}=
 \begin{pmatrix}
 0&0&1\\ 0&0&0\\ 1&0&0
 \end{pmatrix},
\ee
\be
 \lambda_{5}=
 \begin{pmatrix}
 0&0&-i\\ 0&0&0\\ i&0&0
 \end{pmatrix},~~
 \lambda_{6}=
 \begin{pmatrix}
 0&0&0\\ 0&0&1\\ 0&1&0
 \end{pmatrix},~~
 \lambda_{7}=
 \begin{pmatrix}
 0&0&1\\ 0&0&-i\\ 0&i&0
 \end{pmatrix},~~
 \lambda_{8}=
 \frac{1}{\sqrt{3}}
 \begin{pmatrix}
  1&0&0\\0&1&0\\ 0&0&-2
 \end{pmatrix},
\ee
which are the standard $su(3)$ generators \cite{Weigert} and obey the commutation relations:
$[\lambda_{m},\lambda_{n}]=\sum_{s}2if^{mns}\lambda_{s}$,
where the structure constants $f^{mns}$ are completely antisymmetric,
$f^{123}=1$, $f^{147}=f^{165}=f^{246}=f^{345}=f^{376}=\frac{1}{2}$, $f^{458}=f^{678}=\frac{\sqrt{3}}{2}$.
And each two of them are anticommute i.e $\{\lambda_{m},\lambda_{n}\}=0(m\neq n)$.

Let us consider a general qutrit state $\rho$ \cite{Andrei},
\be
\rho=\frac{I+\sqrt{3}\overrightarrow{r}\cdot \overrightarrow{\lambda}}{3},
\ee
where $\overrightarrow{r} \in \mathbf{R}^{8}$ is the Bloch vector of $\rho$ and $\overrightarrow{\lambda}$ is
a formal vector given by the Gell-Mann matrices. For pure qutrit states the Bloch vectors satisfy $|\overrightarrow{r}|=1$, and
for mixed states $|\overrightarrow{r}|< 1$. However, not all Bloch vectors
with $|\overrightarrow{r}|\leq 1$ correspond to valid qutrit states.
For simplicity, we set $r_{2}=r_{3}=r_{5}=r_{7}=
r_{8}=0$, and $r_{1}=a \cos\alpha$, $r_{4}=a\sin\alpha\cos\beta$, $r_{6}=a\sin\alpha\sin\beta$, $|a|\leq 1$.
In this case $\rho$ has the form
\be
\rho=\frac{1}{3}
\begin{pmatrix}
 1&\sqrt{3}a\cos\alpha&\sqrt{3}a\sin\alpha\cos\beta\\
 \sqrt{3}a\cos\alpha&1&\sqrt{3}a\sin\alpha\cos\beta\\
 \sqrt{3}a\sin\alpha\cos\beta&\sqrt{3}a\sin\alpha\sin\beta&1
\end{pmatrix}.
\ee
Then the uncertainty related to the set of observables $\{\lambda_{n}\}^{8}_{n=1}$ has the form,
\be\label{eq12}
\prod^{8}_{n=1}(\varDelta \lambda_{n})^{2}=(\frac{2}{3})^{8}(1-2a^{2}\cos^{2}\alpha)(1-2a^{2}\sin^{2}\alpha+a^{4}\sin^{4}\alpha\sin^{2}2\beta).
\ee

From (\ref{T2}) we have the lower bound of (\ref{eq12}),
\be \label{eq13}
\begin{aligned}
\displaystyle \prod^{8}_{n=1}(\varDelta \lambda_{n})^{2}
&\geq(\frac{1}{3})^{8}(1-2a^{2})[2^{8}(1-2a^{2}\sin^{2}\alpha+a^{4}\sin^{4}\alpha\sin^{2}\beta)\\
&+\frac{1}{8}(-2048+7168a^{2}-6144a^{4}+1359a^{6})\\
\displaystyle \qquad &+\frac{1}{16}a^{2}(4096-6144a^{2}+2385a^{4})\cos2\alpha
+\frac{9}{8}a^{4}(-108+105a^{2})\cos4\alpha\\
\displaystyle \qquad &-\frac{81}{16}a^{5}(\cos6\alpha-32\cos^{2}\alpha\sin^{4}\alpha\cos4\beta)]\\
\displaystyle \qquad &+(\frac{2}{3})^{8}2a^{2}\sin^{2}\alpha(1-2a^{2}\sin^{2}\alpha+a^{4}\sin^{4}\alpha\sin^{2}2\beta).
\end{aligned}
\ee

\begin{figure}[htpb]
\centering\
\includegraphics[width=6.5 cm]{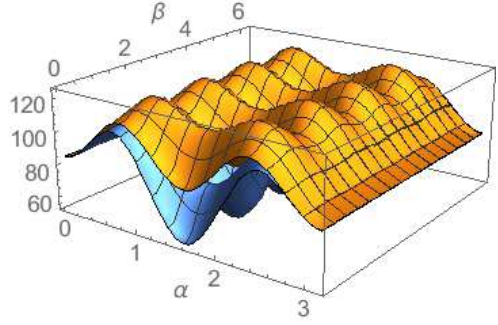}
\caption{{\small (Color online) The uncertainty of observables $\{\lambda_{n}\}^{8}_{n=1}$ in state $\rho$ parameterized by the Bloch vector
$\frac{1}{\sqrt{3}}(\cos\alpha,0,0,\sin\alpha\cos\beta,0,\sin\alpha\sin\beta,0,0)$
and its lower bound. The upper surface is $\Delta=3^{8}\prod^{8}_{n=1}(\Delta \lambda_{n})^{2}$. The lower surface is $3^{8}\times\mathcal{L}_{\ref{eq13}}$, where $\mathcal{L}_{\ref{eq13}}$ is the
lower bound (right side hand of the inequality (\ref{eq13})).}}\label{Fig.2}
\end{figure}

\begin{figure}[htpb]
\centering\
\includegraphics[width=6.5 cm]{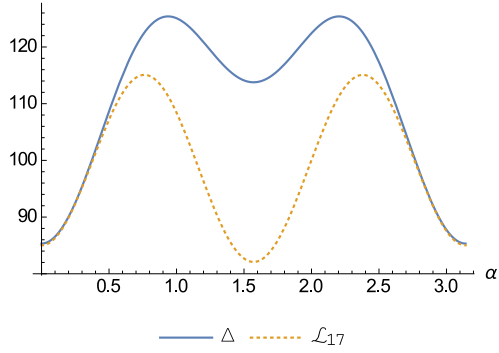}
\caption{{\small (Color online) The uncertainty of observables $\{\lambda_{n}\}^{8}_{n=1}$ in state $\rho$ parameterized by the Bloch vector
$\frac{1}{\sqrt{6}}(\sqrt{2}\cos\alpha,0,0,\pm\sin\alpha,0,\pm\sin\alpha,0,0)$
and its lower bound. The solid line is $\Delta=3^{8}\prod^{8}_{n=1}(\Delta \lambda_{n})^{2}$. The dashed line is $3^{8}\times\mathcal{L}_{\ref{eq13}}$. }}\label{Fig.3}
\end{figure}

When $a^{2}=\frac{1}{2}$, the
equality (\ref{eq13}) holds for all parameters $\alpha$ and $\beta$ corresponding to valid qutrit density matrices. This means that the lower bound (\ref{eq13}) is tight for
$a^{2}=\frac{1}{2}$. For $a^{2}=\frac{1}{3}$, see the Fig. \ref{Fig.2} for the uncertainty relation of these observables.
For explicity, we fix the parameter $\beta$ such that $\sin2\beta=1$, the uncertainty relation is shown by Fig. \ref{Fig.3}.

\medskip
\noindent{\bf Conclusion}

We have investigated the product form uncertainty relations of variances for $n\,(n\geq 3)$ quantum observables.
Tight uncertainty relations satisfied by three observables has been derived explicitly,
which is shown to be better than the ones derived from the strengthened
Heisenberg and the generalized Schr\"{o}dinger uncertainty relations,
and some existing uncertainty relation for three spin-half operators.
Moreover, we also presented a product form uncertainty relation for arbitrary number of observables.
As an example, we first time calculated the uncertainty relation satisfied by the
eight Gell-Mann matrices. Our results have been derived from a class of semi-definite positive
matrices. Other approaches may be also applied to get different types of
product form uncertainty relations for multiple quantum observables.

\medskip
\noindent{\bf Methods}

{\sf Proof of Theorem 1}~To prove the theorem, we first consider the case that all
observables are measured in a pure state $|\psi\ra$. Let us consider a matrix $M$ defined by
\be\nonumber
 M=
 \begin{pmatrix}
  \la\mathbf{A}^{2}\ra &\la\mathbf{B}\mathbf{A}\ra &\la\mathbf{C}\mathbf{A}\ra \\
  \la\mathbf{A}\mathbf{B}\ra &\la\mathbf{B}^{2}\ra &\la\mathbf{C}\mathbf{B}\ra \\
  \la\mathbf{A}\mathbf{C}\ra &\la\mathbf{B}\mathbf{C}\ra &\la\mathbf{C}^{2}\ra \\
 \end{pmatrix},
\ee
where $\la \mathbf{X}\mathbf{Y} \ra=\la \psi|\mathbf{X}\mathbf{Y}|\psi\ra$ for
$\mathbf{X}, \mathbf{Y} =\mathbf{A}, \mathbf{B}, \mathbf{C}$, respectively. For an arbitrary three
dimensional complex vector $\mathtt{x}=(x_{1}, x_{2}, x_{3})\in \mathbb{C}^{3}$, we have
\be
\begin{aligned}\nonumber
\displaystyle \mathtt{x}^{\dagger}M\mathtt{x}&=|x_{1}|^{2}\la \mathbf{A}^{2}\ra+
|x_{2}|^{2}\la \mathbf{B}^{2}\ra+|x_{3}|^{2}\la \mathbf{C}^{2}\ra
+x^{*}_{1}x_{2}\la \mathbf{A}\mathbf{B}\ra+x_{1}x^{*}_{2}\la \mathbf{B}\mathbf{A} \ra\\
\displaystyle&\qquad+x^{*}_{1}x_{3}\la \mathbf{A}\mathbf{C}\ra+x_{1}x^{*}_{3}\la \mathbf{C}\mathbf{A}\ra+
x^{*}_{2}x_{3}\la \mathbf{B}\mathbf{C}\ra+x_{2}x^{*}_{3}\la \mathbf{C}\mathbf{B}\ra\\
\displaystyle&=\la\psi|(x_{1}\mathbf{A}+x_{2}\mathbf{B}+x_{3}\mathbf{C})^{\dagger}
(x_{1}\mathbf{A}+x_{2}\mathbf{B}+x_{3}\mathbf{C})|\psi\ra
\geq 0.
\end{aligned}
\ee
Then for any given mixed state $\rho$ with arbitrary pure state decomposition
$\rho=\sum_{i}p_{i}|\psi_{i}\ra \la \psi_{i}|$, the corresponding matrix $M$ satisfies
$$
\mathtt{x}^{\dagger}M\mathtt{x}=\sum_{i}p_{i}\la\psi_{i}|(x_{1}\mathbf{A}+x_{2}\mathbf{B}+x_{3}\mathbf{C})^{\dagger}
(x_{1}\mathbf{A}+x_{2}\mathbf{B}+x_{3}\mathbf{C})|\psi_{i}\ra
\geq 0.
$$
Therefore $M$ is semi-definite positive for all variance operators $\mathbf{A},\mathbf{B},\mathbf{C}$ and any state $\rho$.
Hence, we have $det(M)\geq 0$, namely,
\be\label{eq4}
\begin{aligned}
\displaystyle & (\Delta\mathbf{A})^{2}(\Delta \mathbf{B})^{2}(\Delta\mathbf{C})^{2} \\
\displaystyle & \geq (\Delta\mathbf{A})^{2}|\la \mathbf{B}\mathbf{C} \ra|^{2} +
(\Delta \mathbf{B})^{2}|\la \mathbf{C}\mathbf{A}\ra|^{2}
+(\Delta \mathbf{C})^{2}|\la\mathbf{A}\mathbf{B}\ra|^{2}\\
\displaystyle &\quad -
2Re\{\la \mathbf{A}\mathbf{B}\ra \la \mathbf{B}\mathbf{C}\ra \la\mathbf{C}\mathbf{A} \ra\}.
\end{aligned}
\ee
By substituting the variance operator $\mathbf{X}=X-\langle X \rangle I$, $X=A, B, C$, into the
above inequality, we obtain the uncertainty relation (\ref{eq3}).
This completes the proof.

\newpage
\bigskip
\noindent{\sf Acknowledgements}

\noindent The work is supported by the NSFC under number 11275131.
Qin acknowledges the fellowship support from the China scholarship council.

\bigskip
\noindent{\sf Author contributions}

\noindent  H.-H.Q, S.-M.F. and X. Li-Jost wrote the main manuscript text. All of the authors reviewed the manuscript.

\bigskip
\noindent{\sf Additional Information}

\noindent Competing Financial Interests: The authors declare no competing financial interests.

\end{document}